\documentclass[12pt]{iopart}
\usepackage{epsfig}
\usepackage{psfig}
\usepackage{floatflt}
\usepackage{subfigure}
\usepackage{setspace}
\usepackage{rotating}

\newcommand{\Xim}{$\Xi^{-}$}
\newcommand{\Xip}{$\bar{\Xi}^{+}$}
\newcommand{\Omm}{$\Omega^{-}$}
\newcommand{\Omp}{$\bar{\Omega}^{+}$}

\newcommand{\2}{$\sqrt{s_{NN}}=200$}

\begin{document}

\title[Multi-strange baryon production at top RHIC energies]{Multi-strange baryon production in Au+Au collisions at top RHIC energy as a probe of bulk properties}

\author{Magali Estienne\dag\ (for the STAR Collaboration)
\footnote[3]{For full author list and acknowledgements, see Appendix ``collaborations'' at the end of this volume.}}

\address{\dag\ SUBATECH, 
4 rue Alfred Kastler, BP 20722, 44307 Nantes Cedex 03, France}
\ead{\mailto{magali.estienne@subatech.in2p3.fr}}

\begin{abstract}
We report STAR preliminary results on multi-strange baryon production in Au+Au collisions at \2 GeV at RHIC. Its implication for the formation of a new state of matter is discussed. The system size dependence on the production of strange baryons is investigated to study the onset of strange quark equilibration in the medium. The nuclear modification factor of $\Lambda$, $\Xi$ and $\Omega$ is also presented. Its suppression at $p_T>3$ GeV/c supports the formation of a dense interacting medium at RHIC. The spectra of multi-strange baryons reveal that within a hydro-inspired model, they may decouple prior than lighter particles and that their flow may be mostly developed at a partonic level. This idea is emphasized by the measurement of the $v_2$ of \Xim$+$\Xip and \Omm$+$\Omp whose behaviour is close to the $\Lambda+\bar{\Lambda}$ baryon elliptic flow in the intermediate $p_T$ region where a constituent quark scaling of $v_2$ is observed.  

\end{abstract}




\section{Introduction}
From Lattice QCD calculations \cite{kars03}, a fast crossover 
between a hadronic matter and a strongly-interacting Quark Gluon Plasma (sQGP) \cite{shur03} should occur at a critical temperature $T_{c}\sim 150-180 MeV$ at vanishing net-baryon density. In this contribution we report on and address some questions pertaining to the physics related to the production of multi-strange baryons in Au+Au collisions at \2 GeV measured by the STAR experiment. Since the mass of s-quark is much higher than that of light quarks and that is close to the critical temperature $T_c$, strangeness is expected to be abundantly produced in a QGP phase at RHIC \cite{rafMu82}. Therefore, the production of strangeness should yield information on the bulk properties from the early stage of the collision to the chemical and kinetic freeze-out. In the first part, their scaling properties with the system size as well as with the number of binary collisions are investigated in order to highlight particle production processes and to look for an onset of strange quark equilibration. In the second part, the transverse collectivity of the collision is studied. Previously, multi-strange baryons have been suggested as being more sensitive to a partonic phase of the collision than to a hadronic phase created after the chemical freeze-out \cite{star04,HSX98,cheng03}. This idea is investigated by looking at their transverse radial flow. As flow builds up through all the collision evolution, its measurement should include both contributions of a partonic part and a hadron one. Multi-strange baryons are of interest to disentangle these two contributions. Furthermore, due to the initial asymmetry of the system in non-central collisions, elliptic flow has also proven to be a well-suited tool for understanding the properties of the early stage of the collisions \cite{Olli92}. A measurement of multi-strange baryon $v_2$ is also presented and discussed. It is directly related to the nuclear modification factor measurement in the intermediate $p_T$ region, 2 GeV/c $< p_T <$ 5 GeV/c to discuss the possible constituent quark degrees of freedom of the matter created before hadronization.

\section{The STAR experiment and the analysis technique}
\label{STAR}
\begin{figure}
\begin{center}
\epsfig{figure=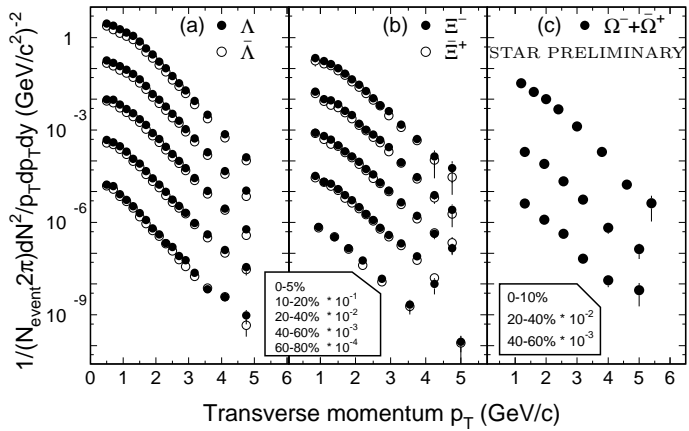,width=9.5cm}
\end{center}
\vspace*{-6.2cm}\hspace*{9.64cm}{\tiny{STAR PRELIMINARY}}
\vspace*{+5.cm}
\caption{\label{fig:spectra}Transverse momentum spectra for $\Lambda$, \Xim and their antiparticles on 5 bins of centrality and on three bins of centrality for $\Omega^{-}+\bar{\Omega}^{+}$ (see text for details). Scale factors are applied to the spectra for clarity. The errors represented are only statistical.}
\end{figure}

The data presented here (1.6$\times$$10^{6}$ minimum-bias and 1.5$\times$$10^{6}$ central events) have been collected by the STAR experiment \cite{Star03}. Inside a magnet delivering a field of 0.5 $T$, the main detector consisted of a cylindrical Time Projection Chamber (TPC) used for track reconstruction and particle identification. Two Zero Degree Calorimeters (ZDCs) measuring neutral spectators along the beam pipe (at forward rapidities) as well as a Central Trigger Barrel (CTB) measuring charged particle multiplicity around mid-rapidity were used as centrality triggers.
Transverse momentum distributions have been measured for $\Lambda$, \Xim, \Omm and their antiparticle. They are corrected for detector acceptance and reconstruction efficiency using the embedding technique where Monte-Carlo particles are embedded into real events. The results for $\Lambda$ ($\bar{\Lambda}$) also include a correction for the week decay of $\Xi$ and $\Omega$. The data cover a domain of rapidity of $|y|<0.75$. Figure ~\ref{fig:spectra} shows the transverse momentum distribution obtained for $\Lambda$, $\bar{\Lambda}$ (a), \Xim, \Xip (b) for 5 bins of centrality : 0-5\%, 10-20\%, 20-40\%, 40-60\% and 60-80\% of the total inelastic cross section. Figure~\ref{fig:spectra} (c) presents the corrected $p_{T}$ distribution for \Omm$+$\Omp for three bins of centrality (0-10\%, 20-40\% and 40-60\%) due to a lack of statistics. The vertical lines represent only statistical errors. The systematic errors which depend on the domain of $p_T$ studied amount 10-15\% of the yields.
The majority of the results discussed in this contribution are extracted from fits to these spectra. The first part of the discussion related to the bulk chemical properties will be derived from the amplitude of the spectra and its scaling with the collision centrality whereas the second part, except for the discussion on $v_{2}$ \cite{v2Cast}, will focus on the shape of the spectra.

\section{Results and discussion}
\subsection{Bulk chemical properties}
\label{par:chemical}
\begin{figure}
\begin{center}
\epsfig{figure=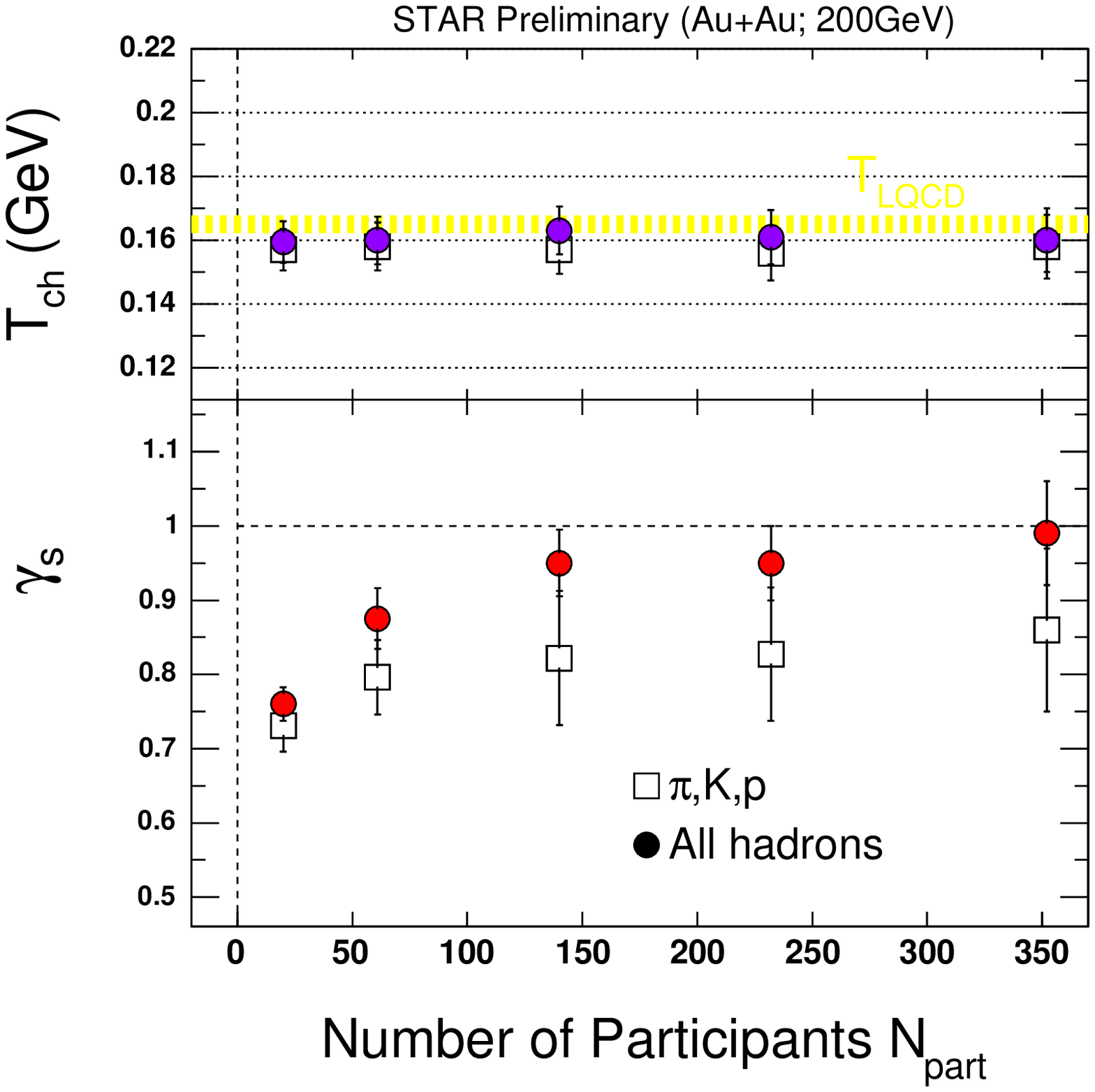,width=7.cm,height=7.2cm}
\epsfig{figure=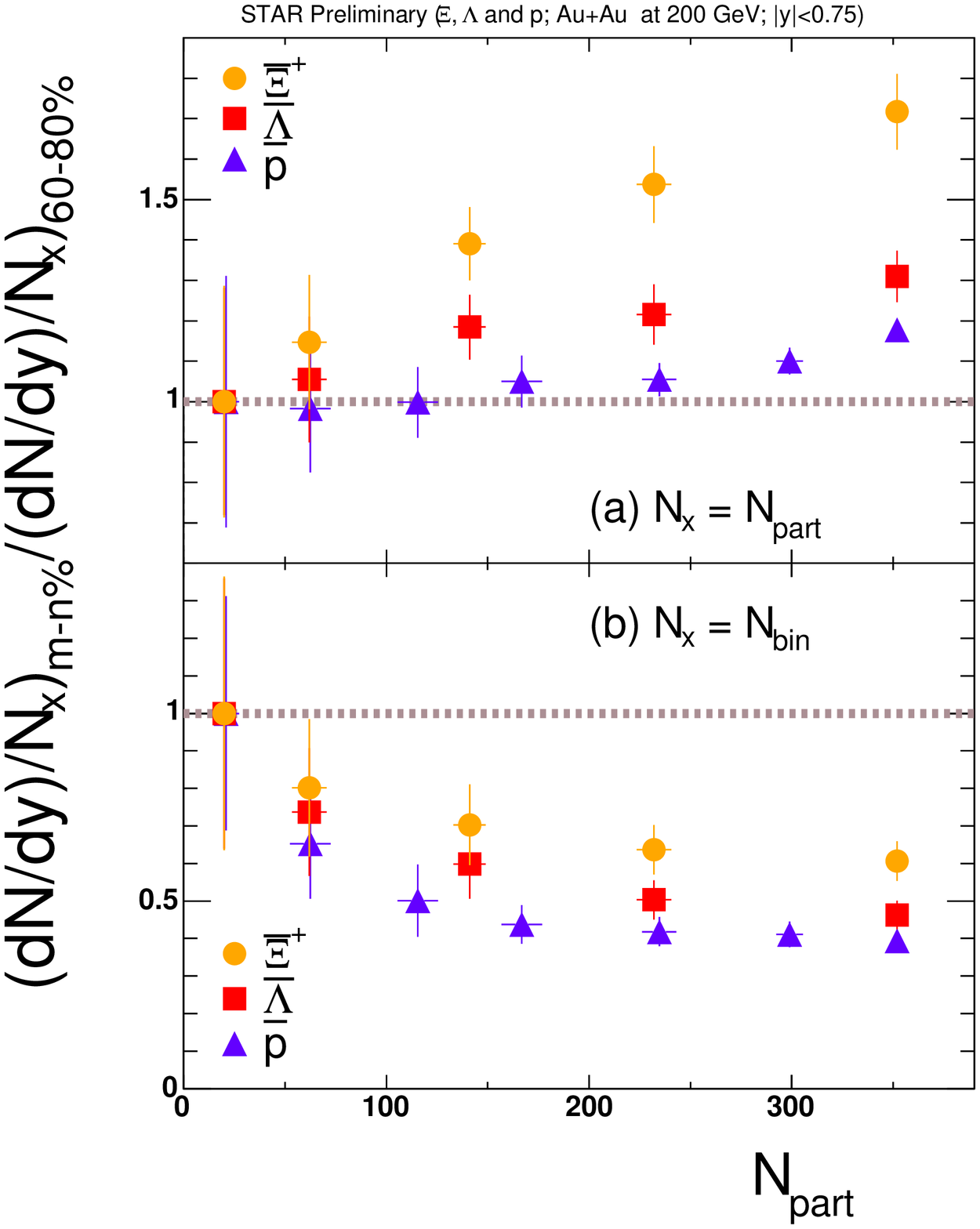,width=6.5cm,height=6.7cm}
\caption{\label{fig:scaling}Left : integrated yields normalized to the number of participants ($N_{part}$) (a) and to the number of binary collisions ($N_{bin}$) (b) as a function of $N_{part}$. The yields are also normalized to the most peripheral one. Right : Chemical temperature $T_{ch}$ and strangeness saturation factor as a function of the collision centrality.}  
\end{center}
\end{figure}

In figure~\ref{fig:scaling} (right), the yields for doubly strange baryons \Xip, the singly strange $\bar{\Lambda}$ and for the non strange baryons $\bar{p}$ normalized to the number of participants (a) and to the number of binary collisions (b) calculated within Glauber model are represented as a function of the collision centrality and with respect to the most peripheral bin. The higher the strange quark content of the baryon, the stronger the increase with $N_{part}$ as it has been suggested previously \cite{Raf03}. The discussion on strangeness enhancement at RHIC and also at lower energies at SPS \cite{caines} is still a matter of debate. One can ask whether full chemical strangeness equilibrium with non-strange hadrons has been achieved in the most central collisions at RHIC and whether hadron production is thermal. Under the assumption that this second point is true, some elements of the answer to the first question can be addressed in the framework of statistical models. They have been successful in describing numerous particle ratios in different system sizes at various energies \cite{review}. In these models, for a particle in a volume $V$, the parameters are the temperature $T$ necessary to achieve a certain average energy density, chemical potentials to constrain the system for baryon number $B$, strangeness $S$ and electrical charge $Q$ conservation. Another parameter of great interest for our purpose is the strangeness phase space occupancy factor, $\gamma_{s}$, expressed as a fugacity. This parameter allows a deviation of strange particles from total chemical equilibrium compared to the lightest $u$ and $d$ quarks. Its evolution as well as $T_{ch}$ evolution with increasing centrality collisions at the top RHIC energy are presented on figure~\ref{fig:scaling} (left). These two parameters have been extracted from fits to the particle ratios \cite{est04} in two cases : (1) only $\pi,K,p$ were considered in the fit (open square), (2) the ratios including all the hadrons (full circle) have been used. Whereas $T_{ch}$ shows no dependence with centrality (close to $T_{c}$ \cite{kars03}), $\gamma_s$ calculation including all hadrons increases from 0.8 to 1.0 within error bars with a saturation reached in the most central collisions. This saturation of $\gamma_s$ close to 1.0 suggests that $s$ quarks are almost chemically equilibrated with the $u$ and $d$ quarks in the most central collisions at RHIC. Between (1) and (2), $\gamma_s$ is not translated upward but is increasing strongly with increasing centrality. This behaviour is related to the inclusion of the multi-strange baryons in the fit whose yields also increase with centrality. The observed saturation should be mostly attributed to the presence in the fit of the light particles abundantly created in the collision. A key to describe the different trends observed in figure~\ref{fig:scaling} (right) could also be found in this model which expresses the particle density as proportional to $\gamma_{s}^{n_s}$, where $n_{s}$ is the number of strange quarks in the particle under consideration. In this description, the $\Lambda$ production should increase linearly with $\gamma_s$ while the $\Xi$ ones are expected to increase as $\gamma_s^2$ and can explain the stronger enhancement observed for the doubly strange baryons compared to the $\Lambda$ ones. 

Figure~\ref{fig:scaling} (right) presents the study of the scaling properties of the yields with the number of participants as well as with the number of binary collisions. The spectra presented in figure~\ref{fig:spectra} demonstrate that the majority of the particles are produced in the low $p_{T}$ region, in a domain presumably dominated by soft physics. In terms of scaling, it implies that the integrated particle yields should be proportional to the volume of the fireball. In a geometrical description, the number of participants is proportional to the initial volume of the system colliding. $N_{part}$ is taken as a reference here. Looking at the distributions for antiparticles in order to compare elements only created during the collisions, a net deviation from the $N_{part}$ scaling is observed for $\bar{\Lambda}$ and most dramatically for \Xip. Looking at figures (b), multi-strange baryons seem to deviate less from a binary scaling than the $\bar{p}$. These observations open different questions and remarks. Is the number of participants completely relevant for multi-strange baryons ? Do they really see this volume where they are created ? Do strange quarks see the same volume as the lightest quarks ? A discussion on this issue was proposed in this conference \cite{caines}. Which processes are at the origin of the strange quark formation ? Are hard processes also relevant in explaining strange quark production ? More pragmatically, is $N_{part}$ a relevant parametrization of the system size for multi-strange baryons as it depends on the way it is calculated \cite{glauber} ?

To go further in the comprehension of hyperon production, we have investigated the dependence of the yields on the number of binary collisions in different domains of $p_{T}$, with the aim of the nuclear modification factor ($R_{CP}$) of strange baryons, defined by comparing the central to peripheral collision yields and scaled to the number of binary collisions for each centrality. In pQCD, the dominant processes involved in particle creation are essentially hard scatterings between nucleons. It provides a reference for $R_{CP}$ in the high $p_{T}$ region where a scaling by $N_{bin}$ is expected. In figure \ref{fig:RcpMB} (right panel), $R_{CP}$ for $\Lambda+\bar{\Lambda}$, \Xim$+$\Xip and \Omm$+$\Omp is represented for the interval 0-5\% vs 40-60\%. The dash line is a guide line for charge hadrons $R_{CP}$. As it has been previously seen for other particles, a scaling by the number of participants appears in the low $p_{T}$ region, in agreement with soft physics processes. In the higher $p_{T}$ region ($p_{T}>3.5$ GeV/c), a suppression of the ratio of multi-strange baryons is also observed instead of a saturation at 1. This observation has found an origin in medium-induced gluon radiation \cite{GVWZ} and seems to highlight the formation at RHIC of a dense interacting medium in the most central Au+Au collisions at top RHIC energy.

\subsection{Transverse collision dynamics}

\begin{figure}
\epsfig{figure=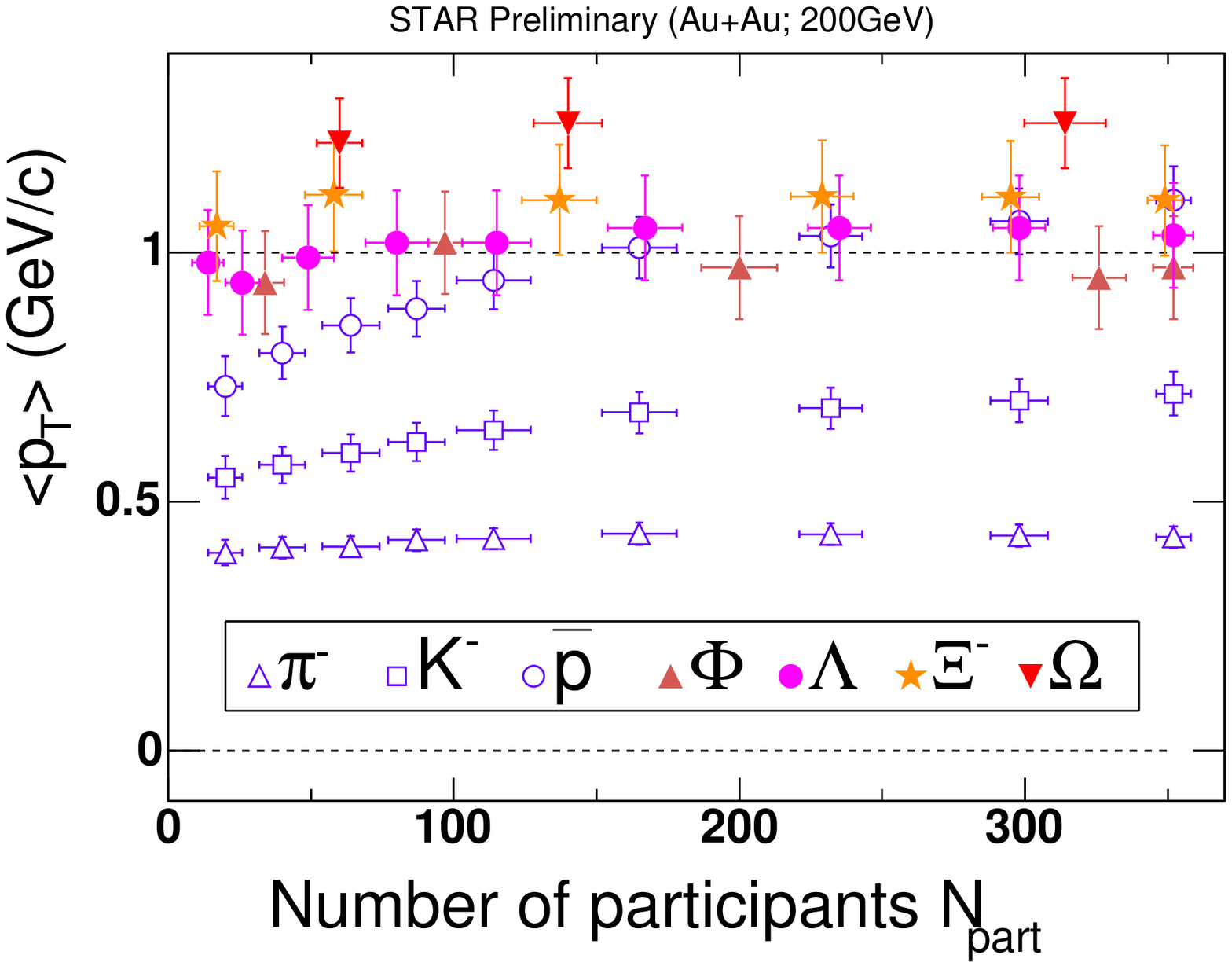,width=8cm}
\epsfig{figure=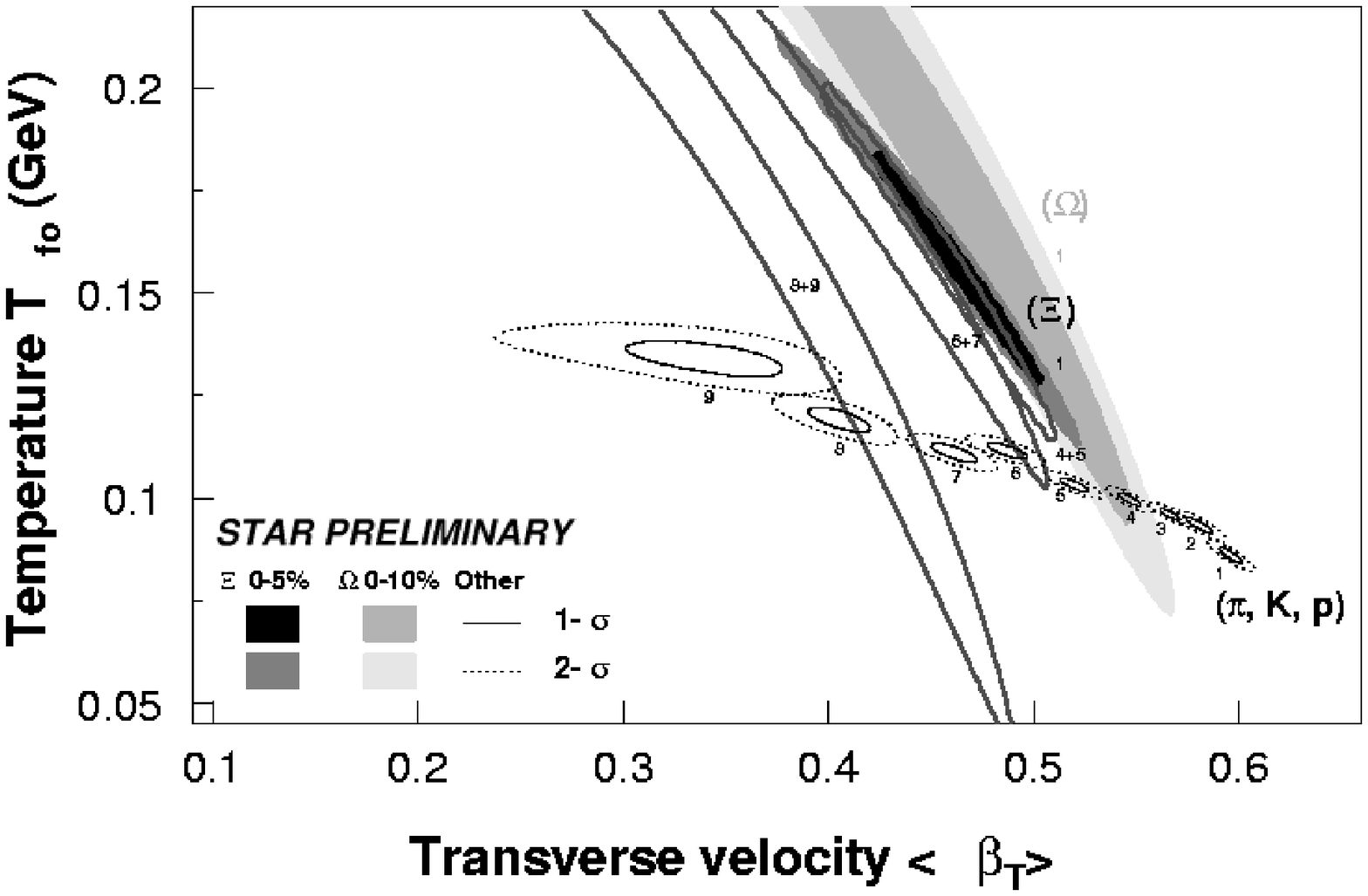,width=8cm,height=6cm}
\caption{\label{fig:collectivity}Left : mean transverse momentum for $\pi, K, p, \Phi, \Lambda, \Xi$ and $\Omega$ as a function of the collision centrality. Right : kinetic freeze-out temperature $T_{kin}$, as a function of the mean collective flow velocity $<\beta_{T}>$, extracted from a hydro-inspired model from $p_T$ distributions. The numbers indicate the centrality selection (see text).}
\end{figure}

In this section, our interest is focused on the study of the transverse collectivity of the collision in order to extract information on its thermal properties as well as the possible formation of a partonic stage during the collision. The mean transverse momentum for various hadrons is presented as a function of centrality in figure~\ref{fig:collectivity}. Whereas the $\pi,K,p$ exhibit behaviours in agreement with a hydro-inspired model \cite{Blas93} - that means an increase of $<p_{T}>$ with centrality and with the particle mass - $<p_{T}>$ of multi-strange baryons show no obvious dependence with centrality within the error bars and their value do not really extend beyond the proton one in the most central collisions. These first observations suggest that multi-strange baryons do not take part in the same collectivity as ($\pi,K,p$) during the collision. 

A hydro-inspired model parametrization known as a ``blastwave fit'' \cite{Blas93}, assuming that all particles are emitted from a thermal expanding source with a transverse flow velocity $<$$\beta_{T}$$>$ at the kinetic freeze-out temperature $T_{kin}$, has been performed on $\pi$, $K$, $p$ spectra together and on $\Xi$ and $\Omega$ separately. A velocity profile $\beta_{T}(r)$=$\beta_{s}(r/R)^{n}$ was used, where $R$ is the radius of the source and $n$ was determined from the fit to the $\pi$, $K$ and $p$ spectra. Its value changed for each centrality bin \cite{est04} and a same value for ($\pi,K,p$) and for $\Xi$ and $\Omega$ has been used for a given range of centrality. Nine bins of centrality indexed from 1 (most central) to 9 (most peripheral) have been extracted for ($\pi$, $K$, $p$), five centrality bins for $\Xi^{-}+\overline{\Xi}^{+}$ and only one (most central) for $\Omega^{-}+\overline{\Omega}^{+}$. The results of the fits are presented in figure~\ref{fig:collectivity} (right panel). The one and two sigma contours are represented for the best fit values ($T_{kin}$,$<$$\beta_{T}$$>$). For the most central bin, we note : 1) there is no overlap of the contours for ($\pi$,$K$,$p$) and ($\Xi$) suggesting that ($\pi$,K,p) take part in the same collective transverse radial flow which is different from that developed by the $\Xi$ ; 2) $\Xi$ seems to kinetically freeze-out at a temperature of $T_{kin}=153MeV\pm20$ MeV, close to $T_{ch}$ (cf.~\ref{par:chemical}) whereas $T_{kin}$ for ($\pi$,$K$,$p$) is approximately $90$ MeV. It suggests that multi-strange particles should have decoupled earlier in the collision close to chemical freeze-out ; 3) furthermore, the fact that the $\pi+\Xi$ interaction cross-section is presumably very small \cite{HSX98,cheng03}, suggests that their flow has been developed prior to chemical freeze-out so prior to the hadronization, probably at a partonic stage. Otherwise, it is corroborated by the fact that the kinetic freeze-out parameters of the multi-strange baryons do not depend on the centrality and that $T_{ch}$ is close to $T_{kin}$. Concerning ($\pi$,$K$,$p$), results show that $T_{ch}>T_{kin}$ and that this temperature difference increases with centrality. It suggests that the time between the two freeze-out is longer for the lightest particles ($\pi$,$K$,$p$) due to rescatterings in the hadron phase while the system is cooling down. In their case, the hadronic part of the flow is not negligeable. 
These results indicate that Au+Au collisions with different initial conditions evolve always to the same chemical freeze-out temperature (\ref{par:chemical}), and then cool down further to a kinetic freeze-out dependent on centrality. Of course the points discussed here have to be taken carefully because the parametrization contains some limitations. The effect of resonances for example on $\pi$ spectra should be seriously studied in the near future. 

\begin{figure}
\epsfig{figure=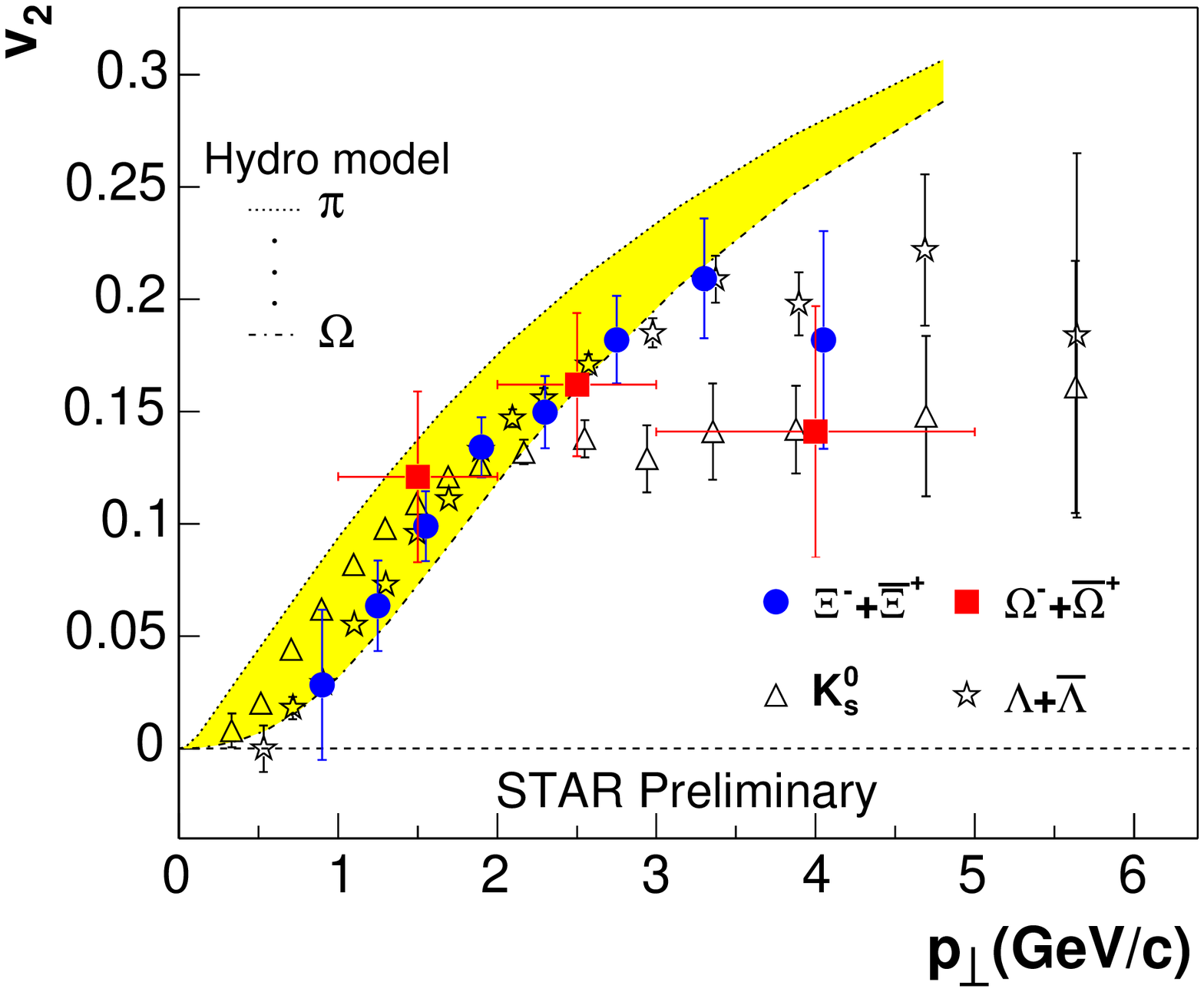,width=7.5cm}
\epsfig{figure=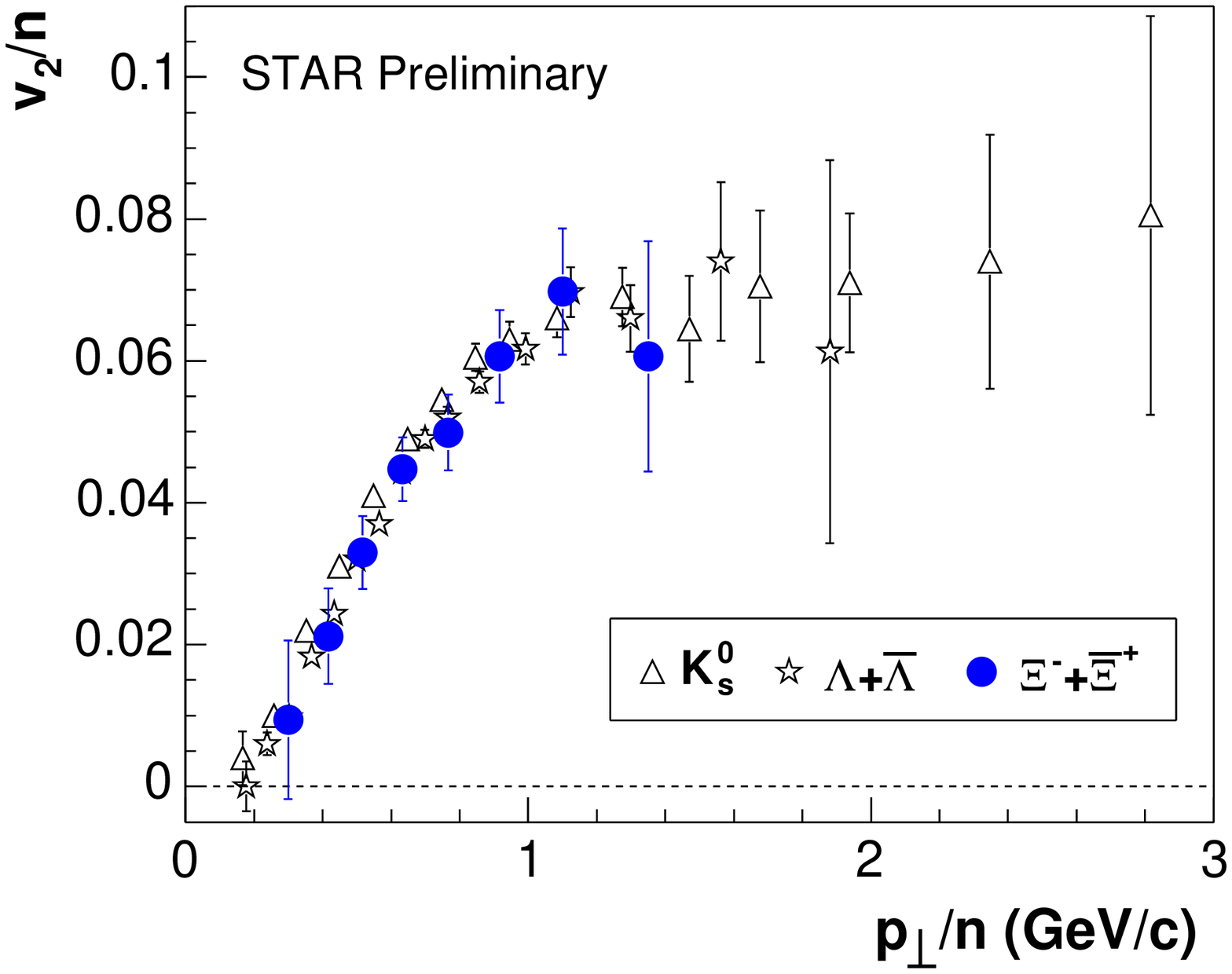,width=7.4cm}
\caption{\label{fig:flow}Left : elliptic flow $v_{2}$ of $K_{s}^{0}$, $\Lambda+\bar{\Lambda}$, \Xim$+$\Xip and \Omm$+$\Omp from \2 GeV Au+Au minimum bias collisions. Hydrodynamic model predictions are shown (colored zone). Right : Elliptic flow $v_{2}$ of  $K_{s}^{0}$, $\Lambda+\bar{\Lambda}$, \Xim$+$\Xip, normalized to the number of constituent quarks ($n$) as a function of $p_{T}/n$ }
\end{figure}
\begin{figure}
\begin{center}
\epsfig{figure=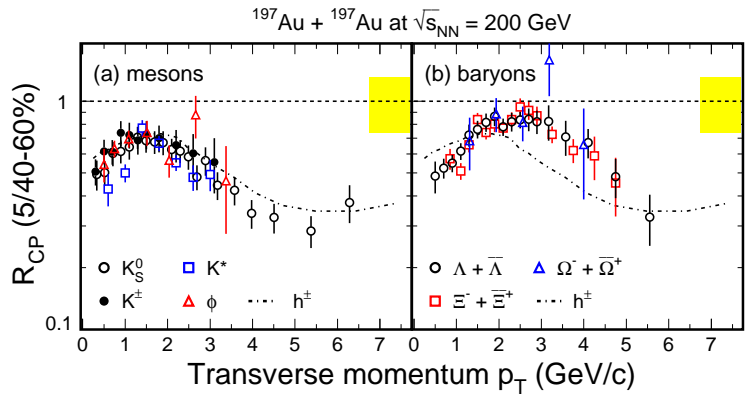,width=10cm}
\vspace*{-0.8cm}
\caption{\label{fig:RcpMB}The nuclear modification factor $R_{CP}$ for mesons (a) and baryons (b) calculated using centrality interval, 0-5\% vs. 40-60\% for \2 GeV Au+Au collisions. The dash line on each panel represents a guide line for all charged particles.}
\end{center}
\vspace*{-13.1cm}\hspace*{4.6cm}{\tiny{STAR PRELIMINARY}}
\vspace*{+12.5cm}
\end{figure}

Hence, this radial flow scenario suggests that for multi-strange baryons, a significant fraction of the transverse flow has been developed probably in a partonic phase of the system so that they should develop elliptic flow. Figure~\ref{fig:flow} shows the measurement of the elliptic flow $v_{2}$ of $\Xi^{-}+\bar{\Xi}^{+}$ and $\Omega^{-}+\bar{\Omega}^{+}$ as a function of $p_{T}$ for the minimum bias data. $v_{2}$ of $K_{s}^{0}$ and $\Lambda+\bar{\Lambda}$ \cite{Soer04} are also presented for comparison. First we observe that the $v_{2}$ of multi-strange baryons is different from zero and seems to follow the same behaviour of the $\Lambda$ $v_{2}$ : the same shape (saturation at a $p_{T}\sim3GeV/c$) and the same amplitude (saturation at $v_{2}\sim20\%$). In the low $p_{T}$ region, the $\Xi$ $v_{2}$ is in agreement with hydrodynamic model calculations \cite{Huo} (colored zone) which predict its mass ordering in this $p_{T}$ region. This model is using an EOS with a phase transition at $T_c=165$ MeV from a partonic system to a hadronic one which freezes-out at $130$ MeV. However, for a $p_{T}>2GeV/c$, the $v_{2}$ deviates from the hydrodynamic model prediction and shows different behaviour for the mesons $K_{s}^{0}$ which saturates around $14\%$ at $p_{T}=2GeV/c$. At the same $p_{T}$ values, a same difference is observed in the nuclear modification factor $R_{CP}$ on figure~\ref{fig:RcpMB} between mesons (left) and baryons (right).  Left panel shows a suppression of the ratio at $p_{T}\sim 2.5 GeV/c$ whereas the suppression for baryons occurs at $p_{T} \sim 3.5 GeV/c$.
It confirms the previously established baryon to meson dependence of the $v_{2}$ and $R_{CP}$ and not a particle mass dependence in the intermediate $p_{T}$ region \cite{Soer04}. This dependence for $R_{CP}$ is supported by the measurement of the $\Phi$ meson, whose mass is close to the $\Lambda$ baryons. $\Phi$ $R_{CP}$ seems to follow the $R_{CP}$ meson shape much more than the baryon one. But the statistical significance of this statement is still low so that a definitive conclusion cannot be drawn. However, this dependence is well explained by quark coalescence or recombination models \cite{Frie03,Greco03,Moln03} in which hadrons are dominantly produced by the coalescence of constituent quarks from a partonic system supporting the idea of a collectivity between partons. These models predict a universal scaling of transverse momentum $p_{T}$ and elliptic flow to the number of constituent quarks ($n$). Such a scaling has been already demonstrated for the mesons $K^{0}_{s}$ and the baryons $\Lambda$ at intermediate $p_{T}$ \cite{Soer04}. Figure~\ref{fig:flow} shows the superposition of the scaled elliptic flows $v_2/n=f(p_{T})/n$ for $K^{0}_{s}$, $\Lambda$ as well as for $\Xi^{-}+\overline{\Xi}^{+}$, supporting that the flow of $s$ quarks is close to that of $u$ and $d$ quarks within error bars.

\section{Conclusion}
A study of multi-strange baryon yields within a statistical model have shown that chemical equilibrium has been achieved in most central collisions at RHIC. The suppression of hadron production at intermediate and high $p_T$ ($p_T>3GeV/c$) central collision demonstrate that the medium created before hadronization is dense and that there are interactions between its constituents. As a matter of fact, the bulk matter created at RHIC exhibits a strong collective expansion. This transverse collectivity has been demonstrated by the measurement of multi-strange baryon radial and elliptic flows. Elliptic flow description within coalescence/recombination model as well as the description of $R_{CP}$ suggest that collectivity has been built up at the partonic level.

\section{Referencing}

\end{document}